\documentclass[aps,preprint]{revtex4-1}

\usepackage{graphicx}
\usepackage{dcolumn}
\usepackage{bm}

\begin{document}
\preprint{}

\title{Asymmetric Energy Transfer Between Coupled Magnetic Vortices by Asymmetric Gyration Amplification}

\author{D. Kumar, S. Barman}
\author{A. Barman} \email{abarman@bose.res.in}
\affiliation{Thematic Unit of Excellence on Nanodevice Technology, Department of Condensed Matter Physics and Material Sciences, S. N. Bose National Centre for Basic Sciences,\\ Block JD, Sector III, Salt Lake, Kolkata 700 098, India.}

\date{\today}

\begin{abstract}

We present a numerical exploration of the possibility of sustained amplification of magnetic vortex gyration by controlling the relative polarities of a coupled vortices in short vortex chains. First, we numerically establish the asymmetry in gyration of a single vortex based on its polarity by use of external magnetic field rotating at the gyrotropic frequency. This phenomena can be used to design logical adapters if vortex core switching is avoided. The criteria to obtain a good gyration amplitude ratio to easily observe true or false output has been examined further. The cases of coupled magnetic vortices and short vortex chains have been studied with different polarity configurations to reveal other desirable aspects of vortex dynamics including, but not limited to, highly efficient signal transfer. These findings are important in applications for information signal processing.

\end{abstract}
\maketitle

\section{Introduction}
Study of inhomogeneous and nontrivial magnetic nanostructures such as magnetic vortices and antivortices is very appealing due to their suggested applications in magnetic data storage, magnetic random access memory devices \cite{Kammerer2011, Jung2011, Kim2012, Jung2012}, magnetic logic devices and information processing devices \cite{Cowburn2002}. Due to the presence of rotational asymmetry they are very attractive candidates for studying the interaction between the local magnetization and externally applied magnetic fields or spin polarized currents \cite{Liu2007, Shibata2003, Shibata2004, Guslienko2006, Lee2008, Metlov2002, Lee2011, Jung2011, Vogel2011, Vogel2012, Sugimoto2011}. In laterally confined ferromagnetic nanodisks, isolated magnetic vortices can be formed, if it is energetically favorable for the magnetization to point in plane and parallel to the edges. In the centre, the magnetization is forced to point out of plane to avoid large angle between the magnetic moments since this will cost large amount of exchange energy. The region with out of plane magnetization is called vortex core and it is only a few nanometers in diameter. The direction of magnetization in the core is termed as polarization and it can have two values $1$ or $-1$ depending upon weather its direction is out of plane or into the plane. Similarly, the rotational sense of the in plane magnetization can be either clockwise (CW) or counter-clockwise (CCW).

Magnetic vortices can be made to gyrate by the application of the magnetic fields and spin polarized currents \cite{Sugimoto2011, VanWaeyenberge2006, Vansteenkiste2009, Curcic2008, Weigand2009, Bolte2008}. External magnetic fields and spin polarized currents couple to the magnetic moments of the vortex core and vortex core moves away from the equilibrium position. In addition to the external forces, the moving vortex core experiences another internal force arising from the demagnetizing field of the non-equilibrium magnetization pattern and this force acts along the perpendicular direction to the vortex core velocity and this leads towards the gyrotropic motion of the vortex core. For large amplitude excitation, the internal force increases nonlinearly and this results into a non linear vortex core gyration, vortex core switching occurs along with creation and annihilation of new vortex and antivortex \cite{Mesler2012, Lee2007}. For small amplitude excitation, the internal force increases linearly and the vortex core motion remains in the linear regime \cite{Thiele1973}.

The gyrotropic vortex core motion is qualitatively described by the Landau-Lifshitz Gilbert equation. In the linear regime, the vortex core equation of motion can be derived from Thiele's equation \cite{Thiele1973}. During the dynamical motion of the vortex core, the demagnetization field at the vortex core point antiparallel to the enlarged domain generated due to the displaced vortex and the time derivative of magnetization vector at the vortex core point into the disk centre or into the opposite direction. A variation in chirality inherently changes the direction of the demagnetization field and therefore does not affect the vortex core motion. Therefore, the vortex core gyration direction is solely controlled by the core polarization. Since the vortex core motion is governed by the demagnetization field, the demagnetization energy determines the vortex core potential. In the linear regime, parabolic potential can be assumed and vortex core can de described in a harmonic oscillator model \cite{Kruger2007}. Therefore magnetostatically coupled vortex gyration in the linear regime can be considered as the coupled forced oscillator. Therefore, one expects mutual energy transfer between the gyrating vortices in the linear regime and a constant phase relation between the gyrating vortices \cite{Barman2010a, Barman2010b}.

The mutual transfer of energy between magnetostatically coupled vortices due to the gyrotropic motion is extremely important for devices for microwave communication \cite{Jung2011}. In this regard, the parameters like the information signal transport rate and energy loss are the key factors in determining the device performance. Vortex gyration transfer rate and energy attenuation coefficients have been calculated by analytical methods \cite{Kim2012} and micromagnetic numerical calculations \cite{Kim2012}. Stimulated vortex gyration based energy transfer between spatially separated dipolar coupled magnetic disks has been observed by time resolved soft x-ray microscopy \cite{Jung2011}.  The rate of energy transfer is determined by the frequency splitting caused by the dipolar interaction between the vortices.

However, the energy transfer efficiency may dependent on several factors such as the frequency of the exciting field pulse as compared to the resonance frequency of the vortex, which is determined by the dimension and aspect ratio of the structure, the amplitude and nature of the exciting pulse, the distance between the two vortices and relative polarization of the two vortices. For the microwave communication devices, one needs to transfer the signal between physically separated but magnetostatically coupled vortices, while only one vortex has been excited for gyration. This can be obtained by application of local magnetic field. However, it has been previously observed that the propagated signal amplitude in the second vortex, is smaller in amplitude if one excites the first vortex in its resonant frequency \cite{Barman2010a, Barman2010b}. It has been observed that the interaction strength between coupled vortices is maximum when the core polarizations of the vortices are opposite \cite{Jung2011}. Higher interaction strength is not the sufficient condition for the higher signal transfer rate. However, to increase the efficiency of the microwave communication devices based on magnetostatically coupled vortices one needs to increase the amplitude of the response and constantly maintain the amplitude. For higher amplitude input, the vortex motion enters into nonlinear regime, vortex core switching occurs via the creation and annihilation new vortices and antivortices. Therefore, in this case it is not possible to increase and maintain a larger amplitude output and a constant phase relation between gyrotropic motion of the both vortices. On the other hand, if the applied signal is of very small amplitude and the frequency is that of the resonance frequency of the vortex, the amplitude of the response gradually increases indicating that the core switching may occur at some point, which is not desirable for the microwave communication devices. One alternative is to use a very small amplitude signal with slight deviation (within 5\% of the resonance frequency) from the resonance frequency (intrinsic) of the vortex. In this letter, we have attempted to obtain a highly efficient signal transfer mechanism, including gyration signal amplification in some cases, by using a non resonant excitation with very small amplitude signal between two magnetostatically coupled vortices with different core polarization.

In Sec.~\ref{sec:method}, we describe the isolated and coupled vortex structures considered here. We also describe the simulation and analysis procedure including a numerical ansatz to achieve a magnetic ground state with desired polarity and chirality \cite{Yakata2011, Jaafar2010, VanWaeyenberge2006, Pigeau2011, Kikuchi2001, Curcic2011, Liu2007, Yamada2007, Jain2012} in a given vortex.

\section{Method \label{sec:method}}

\subsection{Magnetic Vortices}

\begin{figure}[!ht]
\includegraphics[width=8 cm]{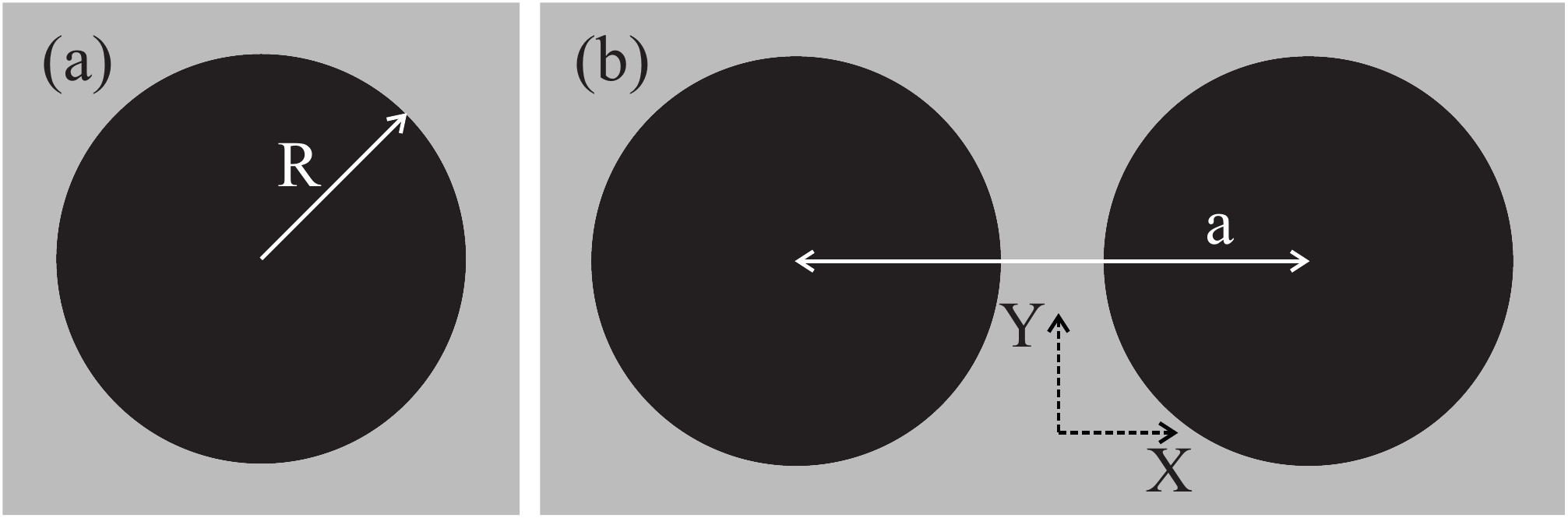}
\caption{Dark regions represent the $40$ nm thick (a) isolated and (b) coupled magnetic vortices each of diameter $2R=200$ nm. The center to center separation in the case of coupled vortices is set to $a=250$ nm.}
\label{fig:vortex}
\end{figure}

A magnetic vortex is formed due to competition between exchange and demagnetizing fields. The use of permalloy (Py: Ni$_{80}$Fe$_{20}$), which has no magneto-crystalline anisotropy, in a $40$ nm thin disk of diameter $2R=200$ nm, which has little shape anisotropy, is expected to support a relatively stable vortex structure. The darker shade in Fig.~\ref{fig:vortex} (a) represents such an isolated vortex. Figure~\ref{fig:vortex} (b) shows a pair of coupled vortices whose center to center distance is $a=250$ nm. A chain of 3 vortices has also been studied with different orientations of polarity. Here too the vortex core, in the absence of perturbation, are at a distance $a$ from each other.

\subsection{Micromagnetic Simulations and Analysis}

Magnetic vortex dynamics was simulated using the finite difference method based Landau-Lifshitz-Gilbert (LLG) ordinary differential equation solver called Object Oriented Micromagnetic Framework (OOMMF). Before the dynamics could be observed, a magnetic ground state needs to be achieved with required vortex core polarity (up or down) an chirality (clockwise: CW or counter-clockwise: CCW). During our simulations, the static magnetic configuration with different combinations of chirality  and polarity were achieved by using a pulse field $H_t=H_0\exp(-t'^2)$. Here, ${\mu_0}{H_0}=1$ T and normalized time $t'=(t-t_0)/(\sqrt{2}\sigma)$. $t_0=75$ ps and $\sigma$ is the standard deviation of this Gaussian pulse in time whose full width at half maximum is $30$ ps. Close to the center of the circular geometries we apply $H_z={\pm}H_t/10$ along the $Z$ axis where the sign controls the core's polarity. If the origin of co-ordinates is brought to the center of the vortex then $X$ and $Y$ components of fields, $H_x$ and $H_y$ that would produce the desired chirality are given below

\begin{eqnarray}
H_x=\mp{H_t}\sin(\theta); \nonumber \\
H_y=\pm{H_t}\cos(\theta).
\label{eq:chirality}
\end{eqnarray}

Here, $\theta=\tan^{-1}(y/x)$ and the upper or lower signs should be chosen for CCW or CW chiralities, respectively. It is to be noted that this pulse signal controlled by $H_t$, dies down quickly while the magnetic ground state is obtained by running the simulation for $40$ ns under a high damping (Gilbert damping constant $\alpha=0.95$ is used in the LLG equation). For Py, we have used saturation magnetization, $M_s=0.8{\times}10^6$ A/m, exchange constant, $A=13{\times}10^{-12}$ J/m and no magneto-crystalline anisotropy. During vortex dynamics simulations we reduce $\alpha$ to a more realistic $0.008$. Magnetization was observed every $10$ ps during dynamics. The cell size used during simulation was $5$ nm $\times 5$ nm $\times 40$ nm.

Upon obtaining the magnetization data from OOMMF, we have chosen to analyze the results by looking at the time evolution of spatial average of normalized $X-$component of magnetization, $\left<m_x\right>$ for each vortex, and its corresponding spectral density. Normalization is done by dividing the $X-$component of magnetization, $M_x$ by $M_s$; such that $m_x=M_x/M_s$. The Hanning window is used on $\left<m_x\right>$ to reduce spectral leakage. The widowed data is then zero padded and Fourier transformed to obtain the required spectral density, $\bar{p}_x$.\cite{Kumar2012} This is plotted on decibel scale as $w_H{\times}20\log_{10}\left|\bar{p}_x\right|$, where a window scaling factor of $w_H=2$ is used for the Hanning window.

\section{Results and Observations \label{sec:ro}}

\subsection{Isolated Magnetic Vortex}

\begin{figure}[!ht]
\includegraphics[width=8.5 cm]{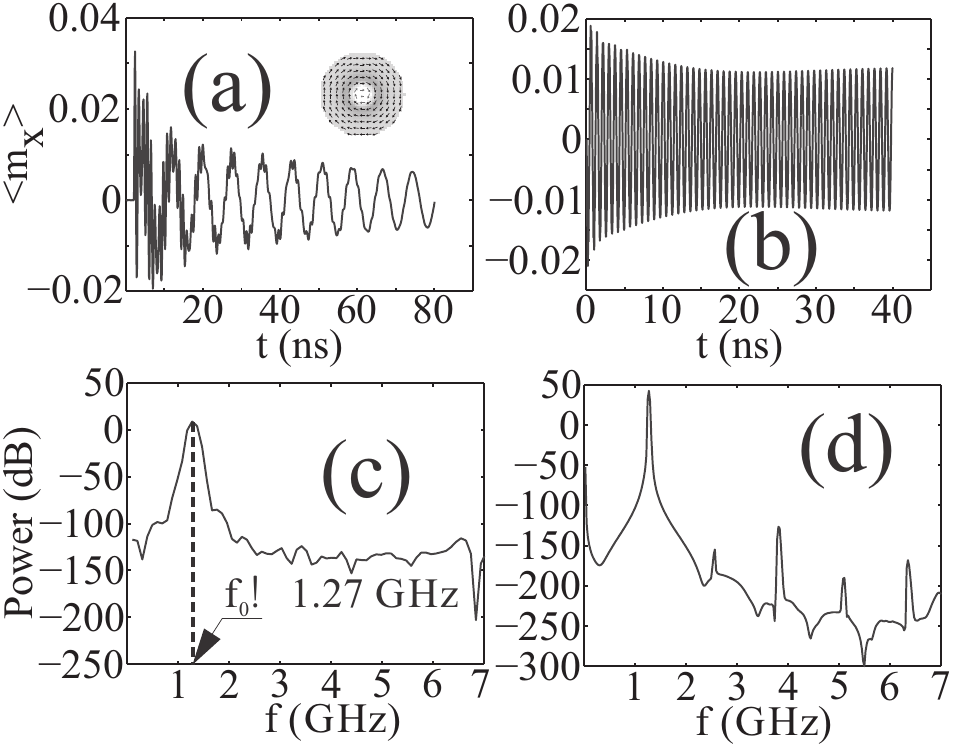}
\caption{Plot of $\left<m_x\right>$ vs. time for (a) signal given by Eq.~\ref{eq:sincS} and (b) signal rotating CCW with frequency $f_0$. The inset in (a) represents the vortex with down polarity in the central white region. Corresponding spectral densities of $\left<m_x\right>$ plotted in (a) and (b) is presented in (c) and (d), respectively. The gyrotropic mode $f_0$ is marked in (c).}
\label{fig:spectrum}
\end{figure}

Before we start to explore the dynamics of magnetic vortices, we first need to obtain the natural frequencies associated with a single isolated vortex. A broadband excitation signal was given to reveal these frequencies. The signal had only $X$-component, $H_{x}^{S}$ which contained power up to $f_{cut}=45$ GHz and depended upon time $t$ as given by:
\begin{eqnarray}
H_{x}^{S}=H_{x}^{0}\frac{\sin(2{\pi}f_{cut}(t-t_0))}{2{\pi}f_{cut}(t-t_0)}.
\label{eq:sincS}
\end{eqnarray}
Here, ${\mu_0}H_{x}^{0}=0.05$ T and $t_0=200$ ps. Figure~\ref{fig:spectrum} (a) shows a plot of $\left<m_x\right>$ vs. time and Fig.~\ref{fig:spectrum} (c) shows the associated spectral density. Only the gyrotropic mode $f_0{\approx}1.2695$ GHz is observed while the higher frequency peaks appear to have been lost during the spatial averaging. The vortex core polarity in the inset of \ref{fig:spectrum} (a) is down, which observes greater gyration amplitude with CW rotation signal.\cite{Lee2008} Thus to observe the smaller peaks at higher frequency, a CCW signal rotating at frequency $f_0$ is chosen with a signal amplitude of $0.5$ mT. Plots of $\left<m_x\right>$ vs. time and the spectral density of the resultant dynamics are presented in Fig.~\ref{fig:spectrum} (b) and (d) where the existence of higher order modes is confirmed.

\begin{figure}[!ht]
\includegraphics[width=16 cm]{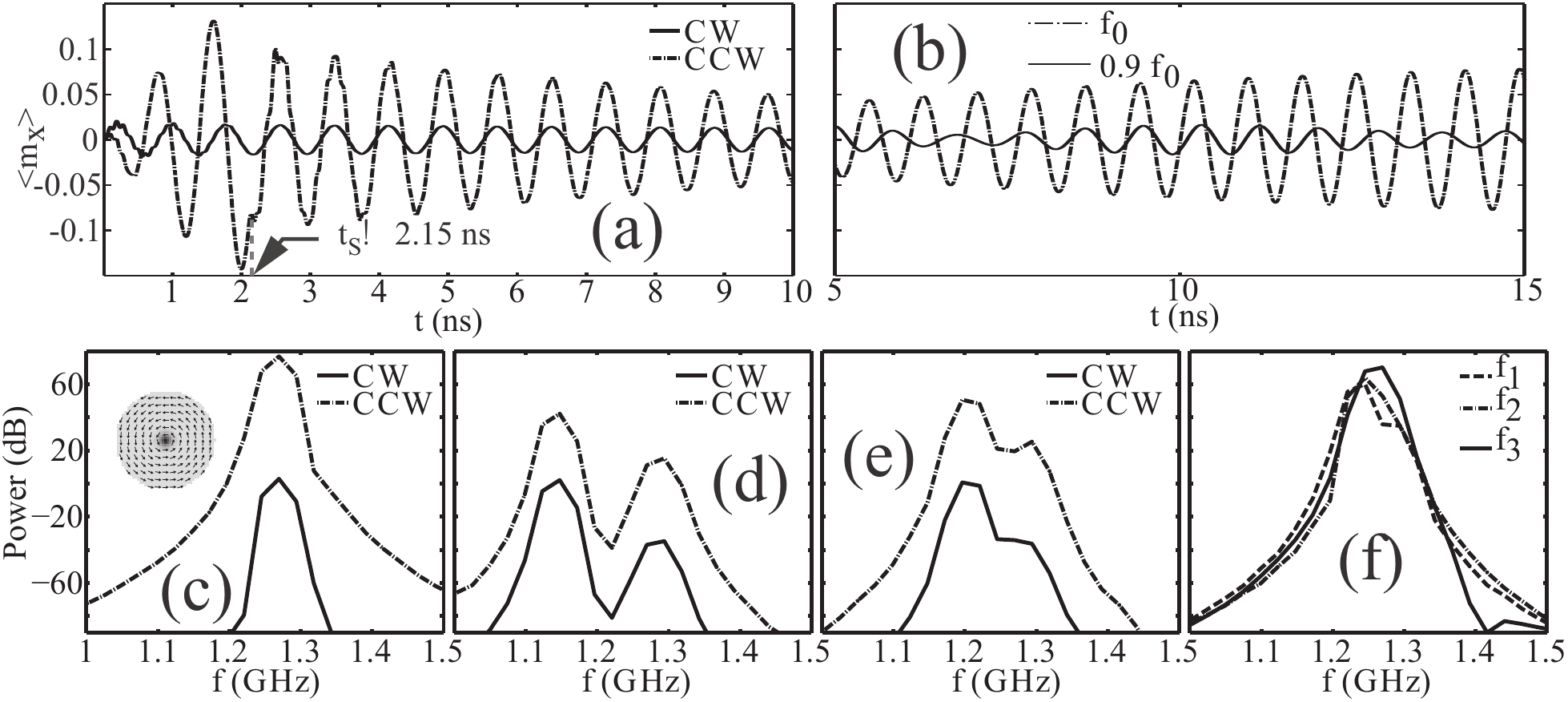}
\caption{(a) Plots of $\left<m_x\right>$ vs. time for signals rotating CW and CCW with a frequency $f_0$ and amplitude $0.5$ mT. (b) Plots of $\left<m_x\right>$ vs. time for signals rotating CCW with a frequencies $f_0$ and $0.9f_0$ and amplitude $0.5$ mT. Spectral densities of vortex dynamics with excitation signal rotating CW and CCW with amplitude $0.5$ mT and frequencies (c) $f_0$, (d) $0.9f_0$ and (e) $0.95f_0$. (f) Spectral density of vortex dynamics with excitation signal rotating CCW with amplitude $0.5$ mT and frequencies $f_1=0.975f_0$, $f_2=0.98f_0$ and $f_3=0.99f_0$. As seen in the inset of (c), the core polarity is up and chirality is CCW.}
\label{fig:comp1}
\end{figure}

Henceforth, we only use signals rotating CW or CCW at frequencies $f_0$ and close to $f_0$. We first use a relatively high signal amplitude of $0.5$ mT on vortex with `up' polarity and CCW chirality (see the inset in Fig.\ref{fig:comp1} (c)). Figure~\ref{fig:comp1} (a) shows the resulting time evolution of $\left<m_x\right>$ for CW and CCW signals rotating at frequency $f_0$. A larger response is achieved in the case of CCW signal. As the gyrating core reaches a certain maximum speed, a polarity switching occurs. In our case, this event occurs at time $t_S \approx 2.15$ ns. This phenomena has been described as potentially useful for data storage in terms of up or down vortex polarity \cite{Kammerer2011}. We find that, if this event is somehow avoided, one can also use this asymmetry in the response for logic operations by attributing the lower and higher resultant gyrations to the binary logical states. Towards that end, we first decided to reduce the signal amplitude to $0.5$ mT. Although, core switching was not observed up till $200$ ns, the convergence happened too slowly ($>100$ ns). As this can be undesirable for logic operations, we next reduced the excitation signal's frequency by $10\%$. Figure~\ref{fig:comp1} (b) shows the time evolution of $\left<m_x\right>$ for signals with amplitude $0.5$ mT and frequencies $f_0$ and $0.9 f_0$. In spite of having a lower amplitude in response, a faster convergence is observed in the later case. Such decrease if amplitude may be undesirable when the logic operation depends upon the capability of distinguishing between higher and lower gyration amplitudes. Figures~\ref{fig:comp1} (c) and (d) show the spectral densities for signals rotating both CW and CCW and frequencies $f_0$ and $0.9 f_0$, respectively. The maxima for CW signals were found to be relatively much less sensitive to the excitation frequency than those for CCW signals. Going from $f_0$ to $0.9 f_0$, the maxima for CW signal drops by 0.774 dB (from $2.907$ dB to $2.133$ dB) while that for CCW signal drops by 35.625 dB (from $76.798$ dB to $41.173$ dB). Like their maxima, the convergence time-scales for CW signals were also found to be largely independent of signal frequency. Hence while considering the trade-off between convergence times and relative signal amplitude one only needs to study the case of CCW signals in order to make a reliable design decision.

We can also see in Fig.~\ref{fig:comp1} (d) that unlike the spectral density of the driven harmonic oscillator, that of the vortex core gyration begins to feature mode-splitting. The beats associated with this split was observed for too long to relate it to the transient solutions of the oscillator. The two peaks seen here are close to $f_0$ and $0.9 f_0$ as best allowed by numerical resolution and spectral leakage. This leads to a beat frequency of about $0.1 f_0$ in the output signal. This is yet another concern, as beats tend to change the output signal's amplitude with time. In Fig.~\ref{fig:comp1} (e), we can see that the signal rotating CW at $0.95 f_0$ produces no beats while that rotating CCW produces beats at a reduced frequency. A gain of power from $41.173$ dB to $50.650$ dB is also observed. Following the trend (of increasing power and reduction in beat frequency below resolution), Fig.~ \ref{fig:comp1} (f) features progressively increasing power maxima for $f_1=0.975f_0$ ($60.143$ dB), $f_2=0.98f_0$ ($63.974$ dB) and $f_3=0.99f_0$ ($70.181$ dB).

\subsection{Coupled Magnetic Vortex}

\begin{figure}[!ht]
\includegraphics[width=8.5 cm]{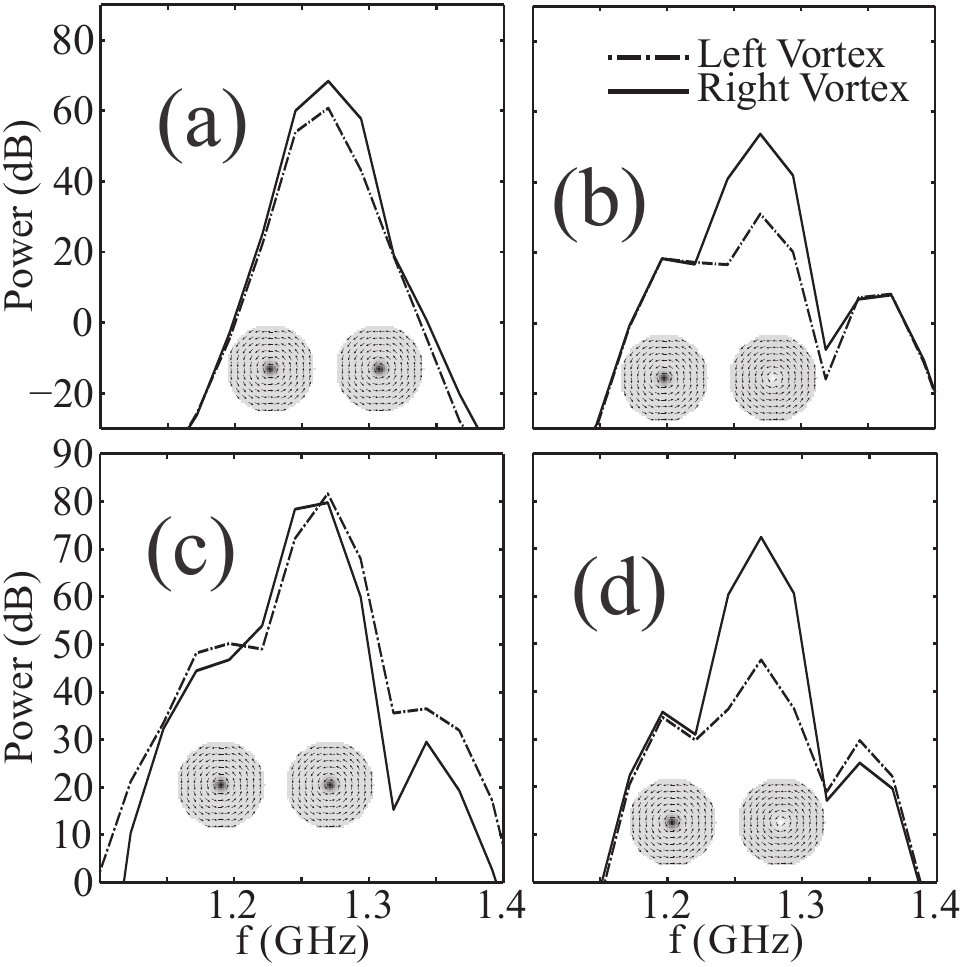}
\caption{Spectral densities of left and right magnetic vortices shown in the insets with ((a) and (c)) similar and ((b) and (d)) opposite polarities. (a) and (b) show the results for a signal amplitude of $0.5$ mT while (c) and (d) show those for an amplitude of $1.5$ mT.}
\label{fig:comp2}
\end{figure}

We have examined the transmission of energy from one vortex to another in terms of their core gyration when excitation is only given to one of them. Like in the previous sub-section, $\left<m_x\right>$ is computed for both the vortices separately. This serves as a indicator of their respective gyrations. The dynamics was examined with all sixteen combinations of polarity and chirality of the two vortices. When a small external bias field is applied, the vortex cores may move up or down along the $Y-$axis. This changes their separation and causes magnetic surface charges to appear on the vortex boundaries; consequently affecting the strength of their magnetostatic coupling \cite{Barman2010}. Hence, in the presence of a bias field, if both the vortices have the same chirality, their coupling will remain relatively unaffected, than when they have different chiralities. This mechanism can used to affect a chirality dependent dynamics and signal transmission. However, in the absence of an external bias, it was quickly understood from the simulated results that chirality does not play any role towards enhancing the asymmetry in dynamics which has been described below. Furthermore, observable changes only appeared to occur between cases with similar and opposite polarities. Mediated by several factors,\cite{Buchanan2005, Belanovsky2012, Zaspel2013, Guslienko2005, Shibata2004} the resonant frequencies of a pair of vortices can differ from that of an isolated vortex. However, in this study, we used an excitation signal rotating at frequency $f_0$ which is applied only on left disk (of vortex pair seen in Fig.~\ref{fig:vortex} (b)). As shown in their insets, Figs.~\ref{fig:comp2} (a) and (c) show the results for the case when both polarities are up ($p_1p_2=1$) and Figs.~\ref{fig:comp2} (b) and (d) show those for the case when left core is up and right core is down ($p_1p_2=-1$). Figs.~\ref{fig:comp2} (a) and (b) correspond to a signal amplitude of $0.5$ mT and Figs.~\ref{fig:comp2} (c) and (d) correspond to a signal amplitude of $1.5$ mT. These spectral densities were calculated after running the dynamics for over $40$ ns, so that any transient vortex core dynamics are suppressed and steady state dynamic solutions appear to be more prominent in the spectrum. 

With increase in the signal amplitude from $0.5$ mT to $1.5$ mT, we see a splitting\cite{Buchanan2007} of the peak. Further, when $p_1p_2=-1$, more power is transmitted and stored in the gyrotropic mode of the vortex which is not being excited directly. These observations testify to the existence of anharmonic and asymmetric dynamics present in the vortex core gyration which cannot be explained by solutions of the Thiele's equation with linear approximations;\cite{Thiele1973, Thiele1974, Huber1982} even if vortex core deformation\cite{Ha2003} and development of dynamic surface charges\cite{Guslienko2002a} are taken into account. This should prompt the community to develop more complex analytical models to better predict such phenomena without the need to do complete simulations.

For signal amplitude of $0.5$ mT and $p_1p_2=1$, the left vortex (which is being excited directly), exhibits $60.72$ dB power in its gyrotropic mode while the right vortex exhibits $68.43$ dB. While for $p_1p_2=-1$, these values become $30.99$ dB and $53.69$ dB, respectively. Thus, about $15$ dB of more power is stored in the gyrotropic mode of the right vortex in the later case ($p_1p_2=-1$). When signal amplitude is increased to $1.5$ mT, these values become $81.66$ dB (left vortex) and $79.76$ dB (right vortex) for $p_1p_2=1$ and $46.67$ dB (left vortex) and $72.45$ dB (right vortex) for $p_1p_2=-1$. Here, about $24$ dB more power is stored in the gyrotropic mode of the right vortex for $p_1p_2=-1$. When $p_1p_2=1$, the power distribution amongst the two vortices is relatively uniform. However, when $p_1p_2=-1$, more power is absorbed in the vortex which is not being excited externally. Also, the discrepancy in power distribution increases with signal amplitude. This shows that the asymmetry based on polarity (regarding where the input power is stored) is enhanced further by signal amplitude. Thus, a logical amplifier can be conceived where input is in the form of external magnetic field and output is read as $\left<m_x\right>$ (e.g., as a Kerr rotation signal).

\subsection{Vortex Chain}

\begin{figure}[!ht]
\includegraphics[width=16 cm]{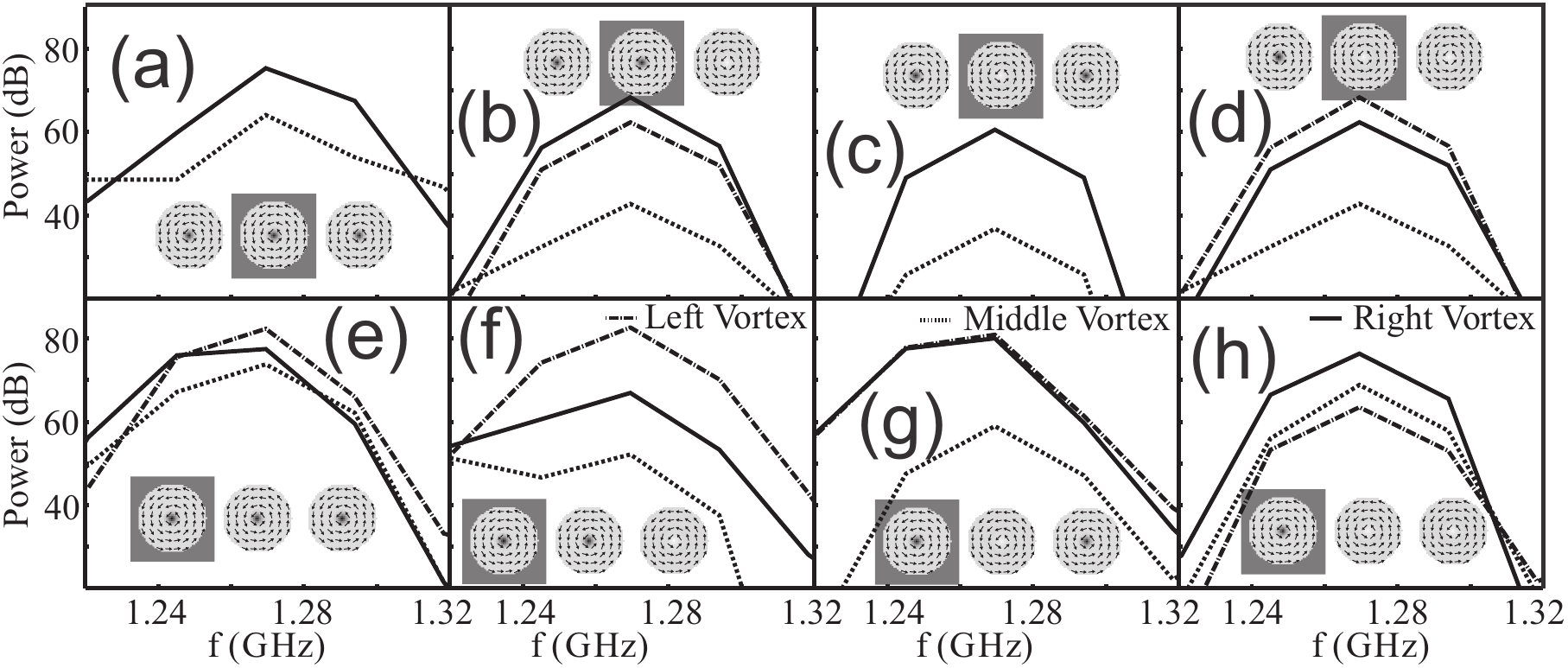}
\caption{Spectral densities of left, middle and right magnetic vortices with ($p_1$, $p_2$ and $p_3$) equaling ($1$, $1$, $1$) for (a) \& (e), ($1$, $1$, $-1$) for (b) \& (f), ($1$, $-1$, $1$) for (c) \& (g) and ($1$, $-1$, $-1$) for (d) \& (h) as shown in the respective insets. The signal is applied only to the vortex with shaded background as shown in the insets. The signal has an amplitude of $1.5$ mT and is rotating CCW except for (c) and (d) where it is rotating CW.}
\label{fig:chain}
\end{figure}

In order to examine the flow of energy stored in the gyrotropic modes of vortices further, we now add another vortex towards the right of the vortex pair shown in Fig.~\ref{fig:vortex} (b) to form a three vortex sequence with polarities (from left to right) $p_1$, $p_2$ and $p_3$, which take values of $1$ or $-1$ depending up or down polarity. We concluded in the previous sub-section that relative polarity is chiefly responsible for the asymmetry in the energy transfer. Hence, here we study only the four cases with $p_1=1$ (up), $p_2={\pm}1$ and $p_3={\pm}1$. Chirality in all cases is CCW. Signal is applied separately to left and middle vortices. The spectral densities for all these eight cases around frequency $f_0$ are shown in Fig.~\ref{fig:chain} (as shown in the insets, the external excitation is given to shaded vortices only). The maximum values of the power at $f_0$ for left, middle and right vortices for all these cases are extracted in Tab.~\ref{tab:comp}, where a bar over $p_i: i\in \left\lbrace 1, 2, 3\right\rbrace$ denotes the vortex to which the rotating signal is applied.

\begin{table}[!ht]
\caption{\label{tab:comp}Power at frequency $f_0$ for different cases shown in Fig.~\ref{fig:chain}.}
\begin{tabular}{|c|c|c|c|c|}
\hline $\mathbf{(p_1, p_2, p_3)}$ & \textbf{Excited Vortex} & \textbf{Left (dB)} & \textbf{Middle (dB)} & \textbf{Right (dB)} \\ 
\hline ($\bar{1}$, $1$, $1$) & Left & 82.32 & 73.79 & 77.46 \\ 
\hline ($1$, $\bar{1}$, $1$) & Middle & 75.33 & 64.08 & 75.34 \\ 
\hline ($\bar{1}$, $1$, $-1$) & Left & 82.64 & 52.16 & 66.93 \\ 
\hline ($1$, $\bar{1}$, $-1$) & Middle & 62.34 & 42.76 & 68.33 \\ 
\hline ($\bar{1}$, $-1$, $1$) & Left & 80.89 & 58.95 & 80 \\ 
\hline ($1$, $-\bar{1}$, $1$) & Middle & 60.55 & 36.78 & 60.57 \\ 
\hline ($\bar{1}$, $-1$, $-1$) & Left & 63.5 & 68.83 & 76.34 \\ 
\hline ($1$, $-\bar{1}$, $-1$) & Middle & 68.33 & 42.77 & 62.37 \\ 
\hline 
\end{tabular}
\end{table}

As seen from Figs.~\ref{fig:chain} (a) and (c), left and right vortices show very similar spectra when excitation is given to the central vortex and $p_1p_3=1$. This symmetry can be broken by manipulating their chiralities under a small steady external field \cite{Barman2010}. However, in the absence of any steady external field, we note (from Tab.~\ref{tab:comp}) that the gyrotropic modes of left and right vortices in the later case (c), exhibit $23.79$ dB more power than that of the middle vortex. This value is $11.26$ dB in the former case (a). It can also be noted that Figs.~\ref{fig:chain} (b) and (d) contain similar information (as the relative polarity arrangement is the same) where $\approx 6$ dB more power is stored in the gyrotropic mode of the vortex with polarity $-p_2$. These observations favor the same conclusion in the previous sub-section that in a given configuration, more power is stored in the vortex with the opposite polarity. Flow of power from first to second and third vortices happens in the arrangements corresponding to the cases shown in Figs.~\ref{fig:chain} (e) to (h). As seen from Tab.~\ref{tab:comp}, power is most evenly distributed (amongst the gyrotropic modes of the three vortices) for ($\bar{1}$, $1$, $1$). In contrast with Fig.~\ref{fig:chain} (e), the power distribution appears to be most uneven in the spectra shown in Fig.~\ref{fig:chain} (f). This suggests that the power distribution is controlled both by the nearest and next-nearest vortices; indicating further non-linearity in the studied dynamics and also hinting at the possibility of controlling the efficiency of power flow with inter-vortex separation \cite{Jung2011}. The difference of power between the corresponding modes of right and left vortices, as seen in Fig.~\ref{fig:chain} (e) is $-4.86$ dB. This value is the least ($-15.71$ dB) in the case ($\bar{1}$, $1$, $-1$), indicating that even in a vortex chain, the most efficient transfer occurs when the neighbouring polarities are opposite. We also observed that the apparent gain in power going from one vortex to another in a vortex pair (previous sub-section) where $p_1p_2=-1$ was not sustained to the third vortex here. Similarly, the significant gain of $12.84$ dB (between right and left most vortices) seen in Fig.~\ref{fig:chain} (h) cannot be expected to remain sustainable over longer chain lengths. Nevertheless, power gain can still be maintained for short vortex chains. Also, transistor like operation can be envisioned with the three vortex sequence considered here when switching $p_2$ from $-1$ (gate open: Fig.~\ref{fig:chain} (h)) to $1$ (gate closed: Fig.~\ref{fig:chain} (f)) changes the difference in power levels of right to left vortex from $12.84$ dB to $-15.71$ dB. ($\bar{1}$, $-1$, $-1$) (Fig.~\ref{fig:chain} (h)) is also the only case considered here where more power is present in the gyrotropic mode of $p_2$ than that of $p_1$. But, the power in $p_2$ is never seen to exceed that in $p_3$.

\section{Conclusions \label{sec:con}}

We numerically examine the polarity dependent asymmetry and non-linearities in vortex dynamics. Cases presented in this paper included isolated vortices and coupled and three vortex sequences. We particularly examine the dynamics for efficiency in power transfer from one vortex to another to arrive at the following conclusions:

\begin{itemize}
\item If polarity switching is avoided by a combination of measures like reduction in signal amplitude and frequency then logical adapters could be designed (in $100$ MHz to GHz range) using the known asymmetric response of isolated vortex to rotating signals. A trade-off between time to reach saturation and ratio of amplitude between logical 0 and 1 outputs needs to be considered during the design.
\item In the case of coupled vortices, when an excitation signal is applied to only one of the vortex then more power is stored in its neighbouring vortex if it has the opposite polarity. A need to develop a non-linear model describing vortex gyration is realized to explain this useful asymmetry.
\item In the absence of such a model \cite{Pokrovskivi1985}, we did further simulations with a sequence of three vortices to confirm our findings. We noticed that most efficient power transfer through a vortex is possible in general when neighbouring vortices had opposite polarity. The apparently anomalous gain in for ($p_1$, $p_2$, $p_3$)$=$($\bar{1}$, $-1$, $-1$) is not deemed sustainable in longer vortex chains. Some of these arrangements can also be used for two input logic gate operators \cite{Jung2012}.
\item A dependence on polarities of both nearest and next-nearest vortices was very evident in the study of different cases of the three vortex sequence. This indicated further presence of non-linear interaction in the dynamics which can be used to affect the efficiency of signal transfer by changing the inter-vortex separation.
\item We also noted that significant gain in terms of gyration mode amplification can be obtained by tuning the relative vortex core polarities of short vortex chains. Thus in the case of three vortex sequence a transistor like operation can be planned where the polarity of the central vortex switches turns the gate on or off causing a relative gain or loss in the power of the right vortex as compared to the left vortex.
\end{itemize}

\begin{acknowledgments}
We acknowledge the financial support from the Department of Science and Technology, Government of India (Grant nos. INT/EC/CMS (24/233552), SR/NM/NS-09/2007), Department of Information Technology, Government of India (Grant no. 1(7)/2010/M\&C).  D. K. would like to acknowledge financial support from CSIR - Senior Research Fellowship (File ID: 09/575/(0090)/2011 EMR-I). 
\end{acknowledgments}

%

\end{document}